\documentclass[aps,prl,floatfix,twocolumn]{revtex4}
\usepackage{bm} 
\usepackage{graphicx}
\usepackage{amsmath}
\usepackage{slashed}
\def\bea{\begin{eqnarray}}
\def\eea{\end{eqnarray}}

\begin{document}
\title{\hfill {\small JLAB-THY-08-858}\\ Is there an $A_y$ problem in low-energy neutron-proton
scattering?}
\author{Franz Gross}
\affiliation{Thomas Jefferson National Accelerator Facility, Newport News, VA 23606\\ and College of William and Mary, Williamsburg, VA 23187}
\author{Alfred Stadler}
\affiliation{Departamento de F\'isica da Universidade de \'Evora, 7000-671 \'Evora, Portugal\\ and Centro de F\'isica Nuclear da Universidade de Lisboa, 1649-003 Lisboa, Portugal}
\begin{abstract}
We calculate $A_y$ in neutron-proton scattering for the interactions models WJC-1 and WJC-2 in the Covariant Spectator Theory. We find that the recent 12 MeV measurements performed at TUNL are in better agreement with our results than with the Nijmegen Phase Shift Analysis of 1993, and after reviewing the low-energy data, conclude that there is no $A_y$ problem in low-energy $np$ scattering.
\end{abstract}

\maketitle

In a recent paper, Braun \textit{et al.}  (designated BR08) report on a new set of very precise measurements of $A_y$ in
elastic neutron-proton scattering at 12.0 MeV neutron lab energy \cite{Bra08}. They compare their
experimental observables to the results produced by the Nijmegen Phase-Shift Analysis \cite{Sto93} of 1993 (PWA93) and by the CD-Bonn potential model \cite{Mac01}, also fitted to the data and producing phase shifts consistent with PWA93. They observe that the predictions of the PWA93 as well as the CD-Bonn differ considerably from the new data and interpret this as evidence for a significant shortcoming of the phase shifts and, consequently, of all potential models fitted to them. In a model study, starting from the CD-Bonn potential, they find that increasing the charged-pion coupling constant $g^2_{\pi^\pm}/4\pi$, while keeping the neutral-pion coupling constant unchanged at $g^2_{\pi^0}/4\pi=13.6$, brings the theoretical predictions into better agreement with the new data.

In this letter, we take up the issues raised in \cite{Bra08} and consider them in light of a recently finished new analysis of \textit{np} scattering within the Covariant Spectator Theory (CST)\cite{Gro07,Gro08}.  As described in detail in Refs.\ \cite{Gro08,Gro92}, the scattering amplitude $M$ for two-nucleon scattering in CST is obtained by solving a covariant integral equation of the form
\begin{equation}
 M=V-VGM 
\end{equation}
where $V$ is the irreducible kernel (playing the role of a potential) and $G$ is the intermediate state propagator.  
The Bethe-Salpeter (BS) equation \cite{Sal51} has a similar structure, but in the BS theory the four-momenta of the particles in intermediate states are subject only to the conservation of total four-momentum $P=p_1+p_2$, so loop integrations over intermediate states are over four independent variables. 
The CST equation maintains four-momentum conservation, but constrains the energy components by placing one of the two intermediate particles on its positive-energy mass shell. Since $P$ is fixed, both intermediate energies become functions of the three-momenta only, and all intermediate loop integrations reduce to three dimensions, as in the non-relativistic theory. In spite of this considerable simplification, the equation remains covariant because the on-mass-shell constraint itself is a covariant condition. This framework has been generalized to other systems and applied successfully to many problems, in particular also to the three-nucleon system \cite{Sta97,Sta97b}.

\begin{figure*}
 \begin{center}
\includegraphics[width=7in]{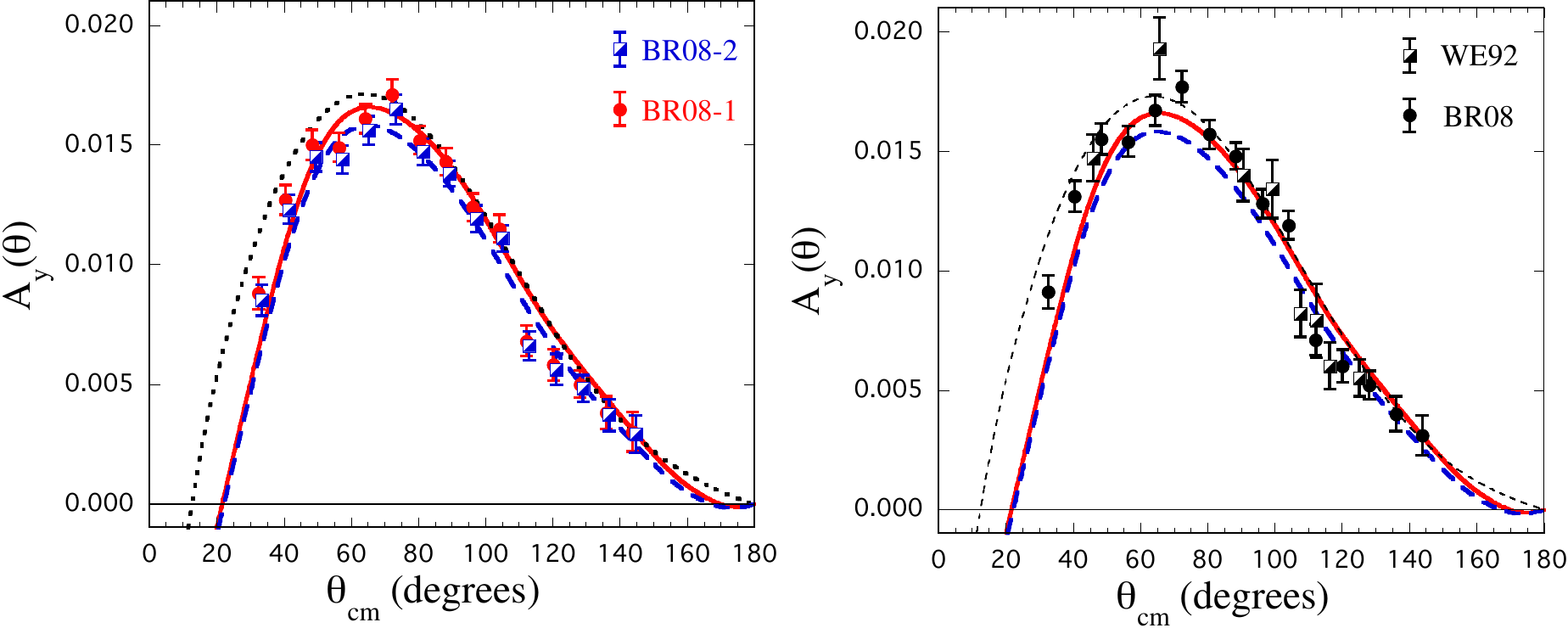}
\end{center}
\caption{\footnotesize\baselineskip=10pt  Neutron-proton analyzing power at 12.0 MeV. Lines are WJC-1 (solid), WJC-2 (dashed) and Nijmegen PWA93 (dotted).  Left panel shows BR08 data scaled by WJC-1 (solid circles) and WJC-2 (half filled squares).  Right panel shows the WE92 \cite{Wei92} data scaled by WJC-1 (half filled squares) and BR08 data scaled by PWA93 (solid circles), }
\label{figure}
\end{figure*} 

\begin{table}
\caption{\footnotesize\baselineskip=10pt  Statistical errors for the data set BR08 of Ref.~\cite{Bra08}.  The systematic error is 1.5\%.}
 \begin{tabular}{lcc|lcc}
 angle$\quad$ & $A_y$ & error(stat) & angle$\quad$ & $A_y$ & error(stat)\\
\hline
  32.6 &  0.00854&0.00066 & 96.3 &  0.01198 & 0.00056\\
  40.5 &  0.01231& 0.00061  & 104.2 &  0.01110 & 0.00055 \\
  48.5 &  0.01451 & 0.00061 & 112.2 &  0.00662 & 0.00061 \\
  56.5  &  0.01443 & 0.00059   & 120.2 &  0.00558 & 0.00064 \\
  64.4  & 0.01560  & 0.00059  &   128.2 &   0.00483 & 0.00056  \\
  72.4  & 0.01659 & 0.00062 & 136.0 &  0.00372 & 0.00067 \\
  80.5 &  0.01470 & 0.00056  & 143.8 &  0.00287 & 0.00079\\
  88.4  & 0.01386 & 0.00053  & & &\\
\hline
 \end{tabular} 
\label{tab:errors}
\end{table}

Our recent fit to the $np$ scattering data \cite{Gro08} uses the CST with a kernel approximated by the sum of one-boson-exchange (OBE) interactions.  Our best model, designated WJC-1, uses 8 bosons with 27 parameters and provides a high precision fit to the 2007 data base ($\chi^2/N_{\rm data}=1.06$).  A new phase shift analysis emerges from this fit, with phase shifts that differ significantly in some cases from the PWA93 analysis.  Another model (WJC-2) was chosen to be as simple as possible, and requires only 6 bosons with 15 parameters.  It does almost as well, with $\chi^2/N_{\rm data}=1.12$.

Before making a detailed assessment, we point out that the errors quoted by BR08 {\it combined systematic and statistical errors together in quadrature\/}.  A better procedure is to treat the systematic normalization error as an independent experimental degree of freedom, so that the $\chi^2$ for the data set would be
\bea
\chi_t^2=\sum_{i=1}^{N}\frac{(o_i/Z-t_i)^2}{(\delta o_i/Z)^2} + \frac{(1-1/Z)^2}{(\delta_{\rm sys}/Z)^2}
\label{syserror}
\eea
where $o_i$ and $t_i$ are the measured and the calculated value of the observable at point $i$, $\delta o_i$ and $\delta_{\rm sys}$ are the statistical errors at point $i$ and the systematic error, $Z$ is a factor [chosen to minimize Eq.~(\ref{syserror})]  by which the data and errors can be divided to correct for the systematic error and improve the agreement with theory, and the last term is the additional contribution to the $\chi^2$ coming from the renormalization of the data.   We contacted the authors of BR08 who told us that the systematic error was about $1.5\%$, and provided us with the original statistical errors, reproduced in Table \ref{tab:errors} \cite{Tor08}.

\begin{figure}
 \begin{center}
\includegraphics[width=3.5in]{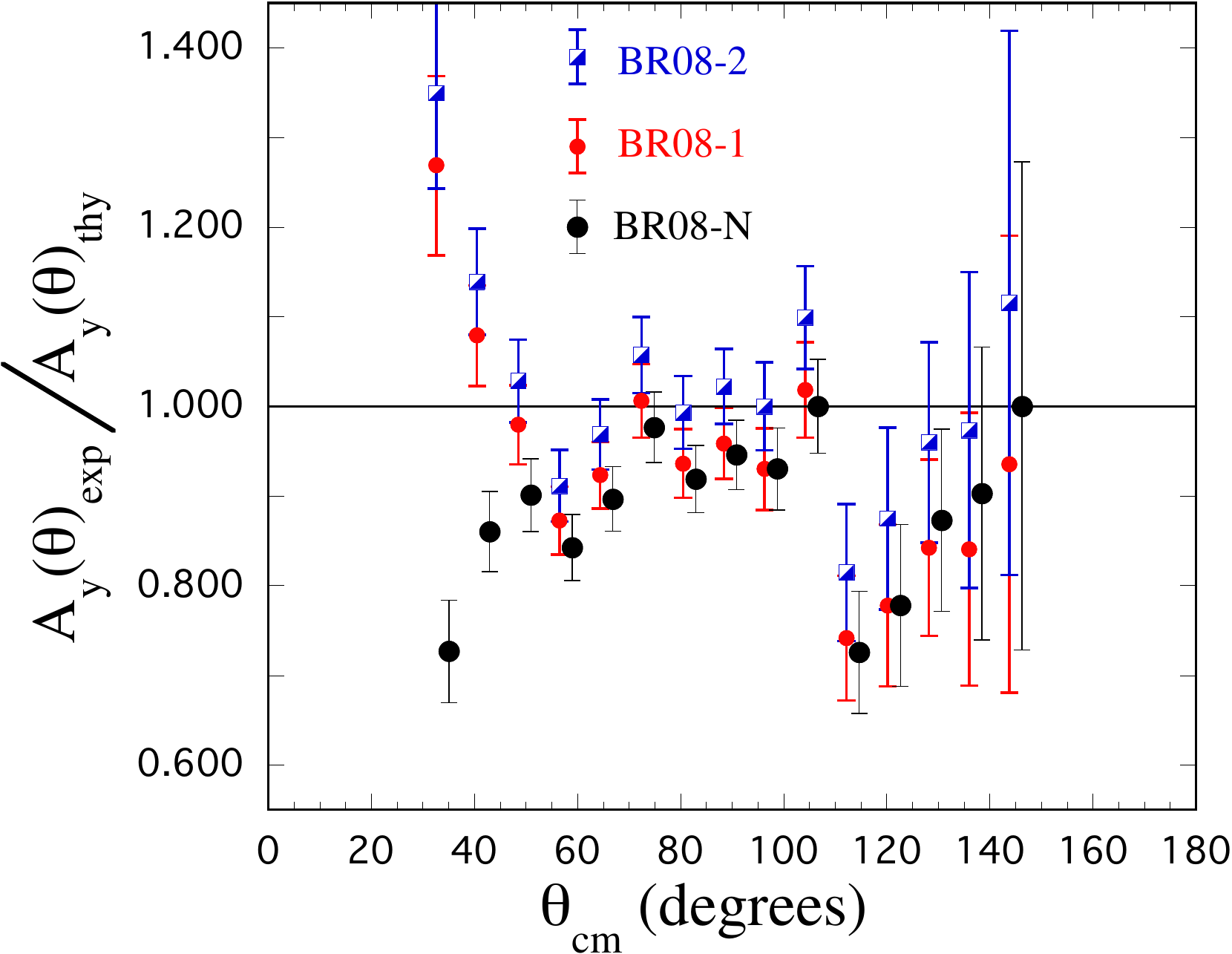}
\end{center}
\caption{\footnotesize\baselineskip=10pt  Ratios of the 12.0 MeV BR08 data for the analyzing power to the predictions of WJC-1 (1), WJC-2 (2), and PWA93 (N).}
\label{figure2}
\end{figure}

Using the new errors given in Table \ref{tab:errors} and the new models (and phase shift analyses), we reach a different assessment of the impact of the BR08 experiment.  Figures~\ref{figure} and \ref{figure2}  present different comparisons of the 12 MeV $A_y$ BR08 measurements with the predictions of the Nijmegen PWA93, and Models WJC-1 and WJC-2. Table \ref{table2} gives the $\chi^2/N_{\rm data}$ for each of these theories and Table \ref{table3} surveys the previous $A_y$ measurements for all laboratory energies less than 20 MeV.  From these figures and tables  we draw the following conclusions:  

\begin{table}
\caption{\footnotesize\baselineskip=10pt The average $\chi^2$ from Ref.\ \cite{Bra08} compared to the average $\chi^2$  obtained by scaling the data and errors in Table \ref{tab:errors} [with the scaling factor $Z$ determined by minimizing Eq.\ (\ref{syserror})].   The scaled WE92 \cite{Wei92} data, also measured at 12 MeV, are shown for comparison. The Nijmegen rejection criteria say that a set with this many data (15 points plus one normalization error) should be retained only if $\chi^2/N_{\rm data}<2.26$.  Note that the BR08 data almost meets this criteria with model WJC-2.  }
 \begin{tabular}{lcccc}
\hline
 & \multicolumn{2}{c}{$\chi^2/N_\mathrm{data}$} &  \\
  Model & BR-Ref.\cite{Bra08} &  $\quad$BR-scaled$\quad$ & $\quad Z$(BR)$\quad$ & WE92 \\
\hline
PWA93(1)\footnote[1]{Our fitting procedure uses the effective range expansion. The numbers shown for PWA93(2) use WJC-1 parameters, which give the best fit to the overall data set at all energies.  Nijmegen does not give $^1S_0$ parameters, but PWA93(1) uses $^3S_1$ parameters taken from Ref.\ \cite{DeS95},  which does not fit the overall data set as well. }\; & 6.621 &  3.849 & 0.9367 & 1.71 \\
PWA93(2)\footnotemark[1]&6.654&3.861&0.9365 &  1.71 \\
WJC-1  & 3.631  & 3.014 & 0.9690 & 1.61 \\
WJC-2  & 2.377 & 2.387 & 1.0029 & 1.16 \\
\hline
 \end{tabular} 
\label{table2}
\end{table}

First, without even looking at the data, observe that the predictions of both of our models WJC-1 and WJC-2 differ substantially from the PWA93 prediction (the theoretical curves shown in both panels of Fig.\ \ref{figure} are identical), expecially at the smaller angles.   This is not unexpected; it is a consequence of the difference between our phase shifts and those of PWA93.  This may have significant impact on other observables as well.

Next, Table \ref{table2} shows that the scale factor $Z$ differs for each model, and hence {\it there is no one ``correct'' data set\/}.  For example, to accommodate PWA93, the data is increased by over 6\% (four standard deviations), giving a large normalization contribution to the $\chi^2$ but lowering the collective effect of the statistical errors on each of the 15 measured points.  This renormalization of the data lowers the $\chi^2/N_{\rm data}$ from about 6.6 to about 3.85.
In the end, even with this renormalization the PWA93 model is the least successful in describing the data.  For comparison, Fig.~\ref{figure} (right panel) and Table \ref{table2} also show previous $A_y$ measurements performed at TUNL at the same energy, designated WE92 \cite{Wei92}. Qualitatively, they appear to agree well with the new measurements, and also with the theoretical models, with considerably smaller $\chi^2/N_{\rm data}$ than BR08, owing mainly to their larger statistical errors.

How are we to interpret these large BR08 $\chi^2$?  Does it mean that all of the theories are wrong?  In this context we must remember that a phase shift analysis gives the {\it most general theoretical description of scattering below the pion production threshold\/}, and while several theoretical models might generate the same set of phase shifts, the {\it best phase shifts\/} provide the best {\it model independent\/} description of the data.  
At low energies, only a few partial waves contribute noticeably to the NN scattering amplitude, and of those only the S-waves are not small. In principle, this leaves the remaining partial waves with large uncertainties. 
However, the low-energy phase shifts are linked to the much better determined phases at higher energies through the requirement of a smooth continuous energy dependence, which limits their uncertainties and relates observables at different energies. While a continuous energy dependence arises automatically when a potential model is used, it is imposed also in a multi-energy phase shift analysis, although in a phenomenological way. 

Thus, any new data set effectively competes with the entire phase shift analysis, which is based in turn on {\it all\/} of the 3788 data currently included in the 2007 data base. If a new data set disagrees with the predictions of the phase shift analysis it also disagrees with all of the other data.

So the BR08 data disagrees with the latest  phase shift analyses (and therefore the entire 2007 $np$ data base), but does this represent a real incompatibility?   Is the 
$\chi^2/N_{\rm data}$ of this data set too high to be tolerated in a quantifiable sense?  The Nijmegen group used a statistical criteria
based on the following consideration: from a given model (in our case a phase shift analysis or a potential model) one calculates the observables contained in a measured dataset. If the model described the observables exactly, in other words, if we had the ``correct'' theory, 
one should obtain $\chi^2/N_{\rm data} \approx 1$ as a result of 
the unavoidable and random statistical errors in the experimental data. This presupposes that each individual data point in the set is normally distributed around the theoretically predicted value with a standard deviation equal to the statistical error. 

One can then ask: based on the theoretical model, and  allowing for random fluctuations of the size given by statistical errors, what is the probability that one will obtain a $\chi^2/N_{\rm data}$ as large as or larger than the actually observed value for the data set? If that probability becomes very small, for instance smaller than some predefined value, one may conclude that the model and the data are not compatible with each other. By the same line of arguments, getting very small values of $\chi^2/N_{\rm data}$ can also be too improbable, leading to a second criteria testing data sets for too low $\chi^2/N_{\rm data}$.

The Nijmegen criteria adopts a critical probability of about 0.27\%, in analogy to the usual ``3$\sigma$-criteria'' for testing individual normally distributed measurements. Accordingly, for $N_{\rm data}=16$ (15 data points plus one common normalization error), the highest admissible value of $\chi^2/N_{\rm data}$ is 2.26. From Table \ref{table2} we see that the BR08 data set is not compatible, in this sense, with any of the models, although it comes very close to WJC-2.

How significant is this incompatibility, and could it be due, in part, to a problem with the BR08 data themselves?  Figure \ref{figure2} shows the ratio of the BR08 $A_y$ measurements to the theory for three different theories. 
There is a rather pronounced disagreement between the data and all models at smaller angles,  but the  ``break'' in the set at about $110^{\rm o}$ may be more significant.   The theories are  in good agreement with each other at these larger angles, yet none of them can reproduce this behavior. 
Eliminating, for example, the two data points at 112$^\circ$ and 120$^\circ$ would already bring the $\chi^2/N_{\rm data}$ of BR08 with respect to both WJC-1 and WJC-2 below the critical value (although not for PWA93). A small increase in the estimated statistical errors would produce a similar result. It is not our intention here to second-guess the data analysis of \cite{Bra08}, but rather to point out that the disagreement of their data with the WJC models is mild and may be partly due to the behavior of the data set itself.

\begin{table}
\caption{\footnotesize\baselineskip=10pt Previous $A_y$ (or $P$) measurements for $E_{\rm lab}\le 20$ MeV.  The tabulated $(\chi^2/N)_1$ are for model  WJC-1; $(\chi^2/N)_2$ are for model  WJC-2.} 
 \begin{tabular}{cccccc}
\hline
$E_{\rm lab}$ & $\qquad$ref$\qquad$ &  $N$ &\;\; $(\chi^2/N)_1$\;\; &\;\; $(\chi^2/N)_2$\;\; & $(\chi^2/N)_{\rm PWA93}$\\
\hline
7.6 & WE92 &  5 & 2.10 & 2.23 & 2.51 \\
10.0& HO88& 13 & 0.76  & 1.25 & 0.82 \\ 
11.0  & MU71  & 1 & 0.04 & 0.13 & 0.03 \\
12.0 & WE92 & 9 & 1.61 & 1.16 & 1.71  \\
13.5 & TO77 & 1 & 0.04 & 0.53 & 0.02 \\
14.1 & BR81 & 11 & 0.36 & 0.42 & 0.37 \\
         & WE92 & 6 & 0.63  & 0.66 & 0.67  \\
14.5 & FI77 & 9 & 0.87  &  0.88 & 0.99 \\
14.8 & TO77 & 1 & 0.35  &  1.67 & 0.41 \\
16.0 & TO77 & 1 & 0.04  &  0.28 & 0.03 \\
         & WE92 & 6 & 1.20 & 1.39 & 1.10 \\
16.2 & GA72 & 3 & 0.17 &  0.18 & 0.18  \\
16.4 & BE62 & 4 & 0.69  &  0.71 & 0.69 \\
         & JO74  & 4 & 0.90  &  1.08 & 0.89 \\
16.8 & MU71 & 1 & 0.03  &  0.07 & 0.02 \\
16.9 & MO74 & 5 & 0.63  &  0.52 & 0.59\\
         & TO88\footnote[1]{This set of forward angle measurements from 51.0 to 143.7 degrees is retained in the fit.} & 12 & 1.30 & 1.30 & 1.33 \\
         & {\it TO88\/}\footnote[2]{This set of backward angle measurements from 136.5 to 166.5 degrees is excluded from the fit {\it because its error is too small\/}.}  &  {\it 5\/} &{\it 0.03\/} & {\it 0.05\/}  & {\it 0.06\/} \\
17.0 & WI84 & 7 & 0.52 & 0.55 &  0.57 \\
18.5 & WE92 & 5 & 0.56  &  0.47 & 0.62\\
19.0 & WI84 & 7 & 0.62  &  0.62 & 0.62  \\
\hline
  &  All  &  111 & 0.85  & 0.90 & 0.90 \\
 \end{tabular} 
\label{table3}
\end{table}

In this context it is  interesting to examine the $np$ database for any other independent evidence of a problem with $A_y$ at low energy.
Table \ref{table3} shows all $A_y$ data for $E_{\rm lab}\le20$ MeV.  All of these sets (except the backward angle TO88 measurements which are excluded because the $\chi^2$ is \textit{too small}) are accepted by the Nijmegen criteria, and all agree well with both WJC models. Thus there is no indication of a problem in the older $A_y$ data.

One might argue that the previous $A_y$ measurements were not as precise as the BR08 data, for instance because they were not corrected for the polarization dependent efficiency of the neutron detectors \cite{Bra08}. Moreover, they have larger errors and are therefore not as restrictive
as the new precise BR08 data. This may be true, but it also means that the larger $\chi^2$ of BR08 is not a result of the older $A_y$ data pulling the fits into a wrong direction, since they could accomodate larger variations in the fits without a prohibitive increase in the overall $\chi^2$. The fits must be dominated by other data.

Our $np$ data base contains 73 data sets below 20 MeV. Only three of them are excluded after applying the Nijmegen criteria: two because their $\chi^2$ is too small, and only a single set, consisting of three total cross section measurements between 0.5 and 2.0 MeV, because of a very large $\chi^2$. So again, we find no hints for any problems in the database. On the contrary, the low-energy data seem to be remarkably consistent with each other and with the phase shift analyses.

This brings us back to the question if the BR08 data set should be excluded from further fits because of its incompatibility with the rest of the data base. From the description given above, it should be clear that, although the Nijmegen criteria leads to a well defined critical value for the maximum tolerated $\chi^2/N_{\rm data}$ of a data set, its choice is also somewhat arbitrary.

\begin{table}
\caption{\footnotesize\baselineskip=10pt Values of the pion coupling constant used in the fits described in this paper.}

 \begin{tabular}{lccc}
\hline
coupling$\qquad$ & WJC-1 &  $\quad$WJC-2$\quad$ &   PWA93 \\
\hline
$g_{\pi^0}^2/4\pi$  & 14.608  & 14.038 & 13.567  \\
$g_{\pi^\pm}^2/4\pi$  & 13.703 & 14.038 & 13.567  \\[2pt]
\hline
 \end{tabular} 
\label{table4}
\end{table}

Instead of worrying too much about whether or not the Nijmegen criteria should be relaxed a bit in order to accomodate the BR08 data set, we decided simply to see if its inclusion in the data base leads to any significant changes in the resulting fits. We found that, because it adds only 16 points (15 plus the normalization error) to a data base of over 3700 points, it had essentially no effect on the fitting process. It is important to realize that the other 3700 points {\it already fix the phases\/} to a very large extent.
We must accept the fact that the BR08 data set has a large $\chi^2$ that cannot be reduced by further fitting.

We are left with some ambiguities.  We {\it cannot\/} conclude, as is suggested in \cite{Bra08}, that the BR08 data point to a need to increase the pion coupling constants. As it turns out, both of the WJC models have larger couplings than advocated by the Nijmegen group (see Table \ref{table4}), but  the BR08 data were not used to obtain these results, and these results are not altered by including the BR08 data in a refit.
Still, this observation may explain why the BR08 data is closer to models WJC.

One {\it cannot\/} escape the fact that it is inappropriate to draw strong conclusions from a single data set. As new data sets are added to the data base, they may slowly change the phase shifts, but any new data set cannot be expected to exert much leverage, and if a new data set disagrees significantly with the phase shift analyses we are rather led to look for problems with the new data set.  Our conclusion at the moment is that the BR08 dataset is marginally consistent with the WJC-2 phases, less consistent with the WJC-1 phases (the best so far), and probably inconsistent with the old PWA93 phases.

{\bf Acknowledgements}: We thank W.~Tornow for helpful correspondence and for providing us with the errors in Table \ref{tab:errors}.  This work is supported by Jefferson Science Associates, LLC under U.S. DOE Contract No.~DE-AC05-06OR23177. A.\ S.\ was supported by FCT under grant No.~POCTI/ISFL/2/275. 

\vspace{-0.1in}

\bibliographystyle{h-elsevier2} 
\bibliography{Ay6a}

\end{document}